\documentclass[aps,twocolumn,amsmath,amssymb,prl]{revtex4}
\usepackage{natbib}
\usepackage{amsmath}
\usepackage{amssymb}
\usepackage{graphicx}

\def\sss{\scriptscriptstyle}
\def\ts{\textstyle}

\def\be{\begin{equation}}
\def\ee{\end{equation}}
\def\ba{\begin{eqnarray}}
\def\ea{\end{eqnarray}}

\newcommand{\mx}{\mbox}
\newcommand{\bm}{\boldmath }

\newcommand{\ax}{$\approx$}

\def\mnras{{MNRAS}}

\def\aap{{A\&A}}
\def\apj{ApJ} 

  \def\prd{Phys.\ Rev.\ D}
  
\begin{document}
\title{The cosmology dependence of weak lensing cluster counts} 
\author{Laura Marian$^{1,2}$, Robert E. Smith$^{3}$, Gary M. Bernstein$^{1}$ \vspace{0.2cm}}
\affiliation{Department of Physics and Astronomy, University of Pennsylvania, Philadelphia, PA 19104, USA $^{1}$ \\ Argelander-Institut f\"{u}r Astronomie, Universit\"{a}t Bonn, Bonn, D-53121, Germany $^{2}$ \\ Institute for Theoretical Physics, University of Z\"{u}rich, Z\"{u}rich, CH 8037, Switzerland$^{3}$}
\email{lmarian@astro.uni-bonn.edu}
\begin{abstract}
We present the main results of a numerical study of weak lensing
cluster counting. We examine the scaling with cosmology of the
projected-density-peak mass function. Our main conclusion is that the
projected-peak and the three-dimensional mass functions scale with
cosmology in an astonishingly close way. This means that, despite 
being derived from a two-dimensional field, the weak lensing cluster
abundance can be used to constrain cosmology in the same way as the
three-dimensional mass function probed by other types of surveys.
\end{abstract}
\maketitle
\mx{\it {Introduction}}---Weak gravitational lensing (WL) has long
been recognized as a powerful cosmological probe. Among the WL
observables are the peaks in shear or convergence maps, which can be
used to detect clusters as points with high signal-to-noise ratio (S/N). The
number of clusters per unit mass per unit volume, also known as the
mass function, $dN/dV/d\log M$ is sensitive to various cosmological
parameters, such as $\Omega_{m}$--the matter density parameter,
$\sigma_{8}$--the normalization of the power spectrum, and $w_{0}$,
$w_{a}$--the dark energy equation of state. Regardless of the way
clusters are detected, the standard approach to constrain cosmology is
to compare their distribution $dN/dV/d\log M$ to mass functions
measured from $N$-body simulations or to the predictions of the
semi-analytical theories such as \cite{1974ApJ...187..425P,
  1999MNRAS.308..119S, 2006ApJ...646..881W, 2008ApJ...688..709T}. In
the case of WL, this procedure has the caveat that shear peaks offer a
two-dimensional statistic, while the above-mentioned mass functions represent a
three-dimensional one: shear peaks can be created by large virialized clusters, but
also by small undetectable clusters, and unvirialized structures
projected along the line of sight. This is known as the projection
effect, a major source of uncertainty in forecasts of the detection
rates and cosmological power of WL cluster-counting surveys.
\par The fundamental assumption used in WL-cluster forecasts
\citep{2006PhRvD..73l3525M, 2004PhRvD..70l3008W} is that the shear-peak
mass function is {\em the same} as the three-dimensional one: given a shear map,
once we have a reliable method to find the peaks, we can use their
distribution to derive cosmological constraints. However, since
line-of-sight projections alter the shear signal of clusters, we do
not know if the abundance of shear peaks depends on cosmology in the
same way as the abundance of virialized clusters.
\par Several studies in the literature have examined the projection
effect. \cite{2001A&A...370..743H, 2003MNRAS.339.1155H} has shown
that cluster mass determination from WL measurements can have errors
of up to 20\% solely due to line-of-sight projections. See also the
related work of \cite{2004PhRvD..70b3008D}. Other numerical studies
\citep{2004MNRAS.350..893H, 2005ApJ...635...60T, 2005NewA...10..676D,
  2005ApJ...624...59H} have compared shear-selected cluster catalogs
to those generated by traditional three-dimensional methods such as the FoF
algorithm \citep{1985ApJ...292..371D}. They have found the relationship
between the measured shear peaks and the expected shear signal
produced by the three-dimensional clusters of their catalogs to be biased and
scattered.

In this Letter, we shall test this fundamental assumption and consider the
following questions: what is the abundance of shear-selected clusters,
i.e. the projected mass function? How does it vary with cosmology?
These two questions mostly determine the cosmological utility of a WL
cluster catalog. We shall address them numerically. Pioneering work on
this topic has been done by \cite{1999MNRAS.302..821K,
  1999A&A...351..815R}. They compare several cosmological models and
use the aperture mass technique of \cite{1996MNRAS.283..837S} to
conclude that the shear-derived cluster abundance is in reasonable
agreement with the Press-Schechter theory.
The conclusions which we draw are based on the analysis
of a large ensemble of 32 $N$-body simulations, sampling four different
cosmological models with eight realizations per model. Our results
therefore have more statistical power than previous studies, since we
have a significantly larger sample volume and also a larger number of
independent realizations per model. This latter fact enables us to
place robust ensemble-averaged errors on our results.

In a WL map, the line-of-sight structures which contribute to a
cluster's signal can be nearby the cluster, in which case they are
correlated with it, or they can be at large distances adding
accidentally to the signal of the cluster. Therefore, we can
distinguish between {\em correlated} and {\em uncorrelated}
projections. Our focus here is {\em only} on correlated projections:
we divide our simulation volume into slabs with thickness in the range
where the cluster correlation function is significant. We defer
estimating the impact of chance projections to future work, as they
require very large simulation volumes. In addition, previous studies
have found chance projections to be of small significance, e.g. see
\cite{2004MNRAS.350..893H, 2005NewA...10..676D,
  2006PhRvD..73l3525M}. From the simulations, we measure the peaks in
the projected density field at a redshift of interest for weak lensing
surveys, $z=0.3$. For the same redshift, we then measure the three-dimensional mass
function of the halos and compare to the projected-peak mass function
and also to the theoretical prediction of
\cite{1999MNRAS.308..119S}. The projected density field is not a WL
observable. However, since we do work in thin slabs, it is equal to
the convergence up to a proportionality constant. This is a simplified
case, but should the projected-peak mass function thus measured behave
very differently than the prediction of \cite{1999MNRAS.308..119S}, we
expect that in a more realistic scenario, where convergence or shear
peaks are measured, the discrepancy would only increase. We would also
like to separate very cleanly the impact on the two-dimensional mass function of
the correlated projection bias from biases induced by other WL survey
systematics, such as shape noise.
\par \mx{\it{Methodology}}---We have used Gadget2
\citep{2005MNRAS.364.1105S} to generate simulations with $400^{3}$
particles in a volume of $512^{3}$ $\rm({Mpc/h)^{3}}$; the simulations
were started at $z=50$, and the initial conditions were generated with
2LPT \citep{2006MNRAS.373..369C}. For each of the eight realizations, the
initial conditions for the four cosmologies are the same. This matching
of initial conditions will minimize the cosmic variance on the
comparison of mass functions in different cosmologies.

Given our focus on correlated projections, we divide each simulation
cube into 10 slabs normal to the chosen line of sight. Each slab has a
thickness of \ax 50 Mpc/h, which is the range where the cluster
correlation function is significant. We convolve the projected matter
density in these slabs with an optimal filter, to find the density
peaks. The filter we use is similar to that described in
\cite{2006PhRvD..73l3525M}. To every point in the filtered density
field map we associate a mass given by:
\be
M(\mx{\bm $x_{0}$})=\int d^{2}x \, W(\mx{\bm $x_{0}$}-\mx{\bm $x$})
\Sigma(\mx{\bm $x$}),
\label{eq:fil1}
\ee
where $W$ is our filter and $\Sigma$ denotes the projected density of
dark matter. $W$ is an optimal filter, i.e. it maximizes the
$\rm{S/N}$. $W$ is tuned to best recover halos with Navarro-Frenk-White
(NFW) density profiles, but with masses defined by Sheth-Tormen (ST):
if a peak identified in the projected density map corresponds to such
a cluster, then the filter will return the ST mass of that cluster at
the position of the peak. ST virial radii, concentrations, and masses
are larger than the NFW ones. We choose to normalize our filter for ST
masses over NFW ones in order to get a better match with the
three-dimensional halos of our simulations, which were identified with
the FoF algorithm. The FoF halo mass function fits the ST mass
function with an accuracy of $\approx 10\%-20\%$. The filter satisfying
these requirements is given by the expression:
\be
W(\mx{\bm $x$})=M_{\sss ST}\, \frac{\Sigma_{\sss ST}(\mx{\bm $x$})}
{\int d \mx{\bm $x$}\left|\Sigma_{\sss ST}(\mx{\bm $x$})\right|^{2}},
\label{eq:fil2}
\ee
where we have used the truncated ST projected density:
$
\Sigma_{\sss ST}(x)=\ts{r_{s}\,\delta_{c}^{\sss ST}\,\rho_{m}}\,f_{\sss
ST}(x).
$
 $r_{s}$ is the scale radius of the cluster, $\rho_{m}$ is the matter
density at the redshift of the halo, $\delta_{c}^{\sss ST}$ is the
characteristic overdensity of the profile, and $f_{\sss ST}$ is a
function which depends on cosmology only through the concentration
parameter. For its expression, see, for instance,
\cite{2004MNRAS.350..893H}. 

Having filtered the two-dimensional density map, we then apply a rigorous
algorithm to find the peaks. The novelty of our peak-finding algorithm
is that we filter the density field recursively, with filters of
decreasing mass. The highest mass peaks are identified first; smaller
peaks subsequently found at the same location or within the virial
radius of a higher peak are discarded. Thus the ``halos in halos''
problem is nicely solved, e.g. see
\cite{2004MNRAS.350..893H}. Moreover, we understand the variation with
cosmology of our filtered output, which is crucial for establishing
the cosmology scaling of the lensing mass function. Finally, our
filter has a very clear physical meaning, i.e. it returns the mass of
a halo. All these features distinguish our peak-finding technique from
other WL cluster studies, which use Gaussian filters of fixed size,
see, for instance, \cite{2004MNRAS.350..893H, 2005ApJ...635...60T}. Full
details of our projected-cluster-finding algorithm, the numerical
simulations used, and an analytic model for predicting the projection
noise contamination are described in a companion paper \cite{2009M}.
\begin{figure*}
\centering
\includegraphics[scale=0.65]{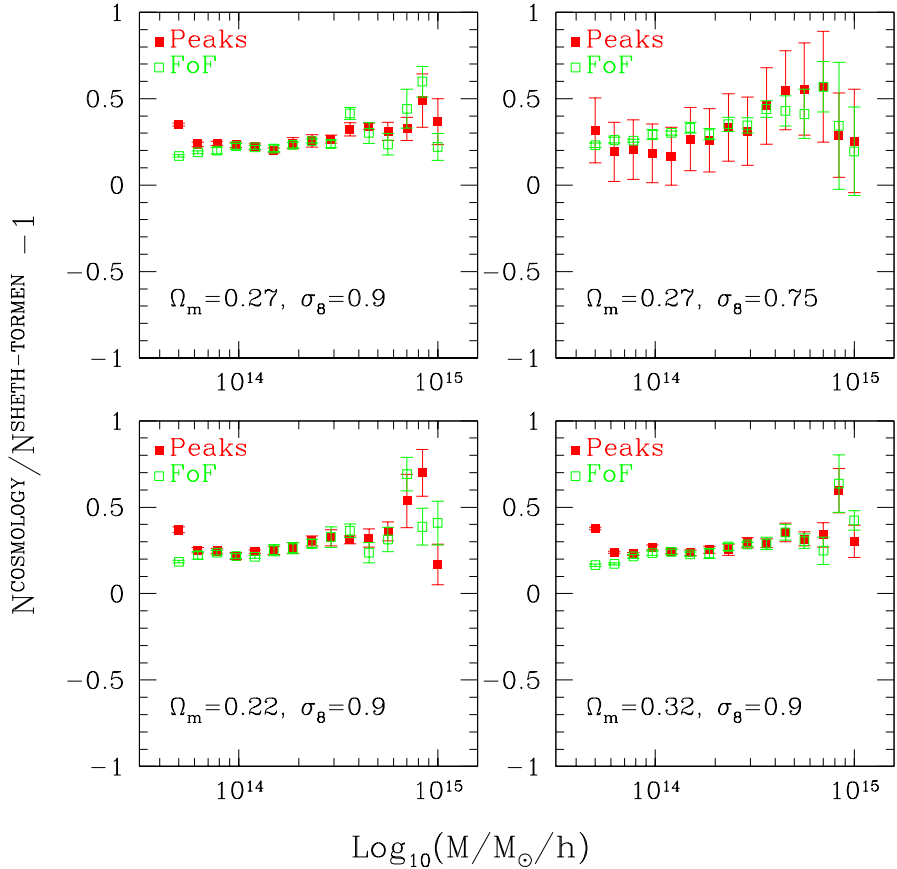}
\caption{The projected-peak (solid squares) and the FoF (empty
squares) mass functions, scaled by the ST theory for the four
cosmologies of our simulations. The error bars represent errors on the
mean of the eight realizations of each cosmology.}
\label{fig:mf_abs}
\end{figure*}
\par \mx{\it {Results}}---In Figure~\ref{fig:mf_abs} we present the
projected-peak and the three-dimensional halo mass functions, measured from our
simulations and ratioed with respect to the predictions from the ST
theory. The cosmologies of the simulations assume a flat universe,
with dark energy in the form of a cosmological constant. We chose our
fiducial model to have $\Omega_{m}=0.27, \, \sigma_{8}=0.9$. We shall
refer to the other three models as the variational cosmologies,
because they simply vary the matter density parameter and the
amplitude of the power spectrum around the fiducial values:
$\Omega_{m}=\{0.22,\, 0.32,\, 0.27\}$, and respectively
$\sigma_{8}=\{0.9,\, 0.9,\, 0.75\}$. Thus we are directly
testing the cosmological dependence of the lensing mass function.

Each panel in Figure~\ref{fig:mf_abs} corresponds to one of these
models. The projected density peaks are depicted by the red solid
rectangles, while the three-dimensional halos are the green empty
rectangles. The error bars have been computed as errors on the mean of
the eight realizations of each cosmology. For all four cosmological
models, we find the mass function of three-dimensional halos
comparable to that of projected peaks. The difference between the two
mass functions is due to both projection effects and also to the
efficiency with which we recover the halos from the projected density
field. If we were to consider only the projection effects, we would
expect the lensing mass function to be higher than the
three-dimensional one, which it is not the case for most of the mass
bins shown in Figure~\ref{fig:mf_abs}. This is mainly due to the fact
that we compare peaks found with a spherical-overdensity (SO) filter
to FoF halos. FoF and SO masses are known to differ, because the FoF
algorithm links together small halos that are close to each other,
whereas the SO halo finder considers them as separate objects.  For a
rigorous analysis of this issue, see \cite{2008ApJ...688..709T} and
references therein. \cite{2001ApJ...547..560M} and
\cite{2004MNRAS.350.1038C} have thoroughly investigated the
correspondence between three-dimensional and two-dimensional masses of
individual clusters in the presence of correlated projections and have
found significant scatter between the two. The former concluded that
cluster masses are overestimated due to line-of-sight projections,
while the latter study attributed the scatter mostly to the triaxial
nature of halos. We also found that when filtering isolated FoF halos
(i.e. only the particles from a single halo), in more than 90\% of
cases we recover the mass to be smaller or equal to the FoF mass,
which explains the trend displayed in Figure~\ref{fig:mf_abs}. It is
however difficult to compare our results to these previous works, as
the mean ratio of two-dimensional and three-dimensional masses depends
on the mass definition and filter used. We also did not consider mass
estimates averaged over several line-of-sights, as done by
\cite{2001ApJ...547..560M} and \cite{2004MNRAS.350.1038C}, but
analyzed a much larger number of clusters.
\begin{figure}
\centering
\includegraphics[scale=0.45]{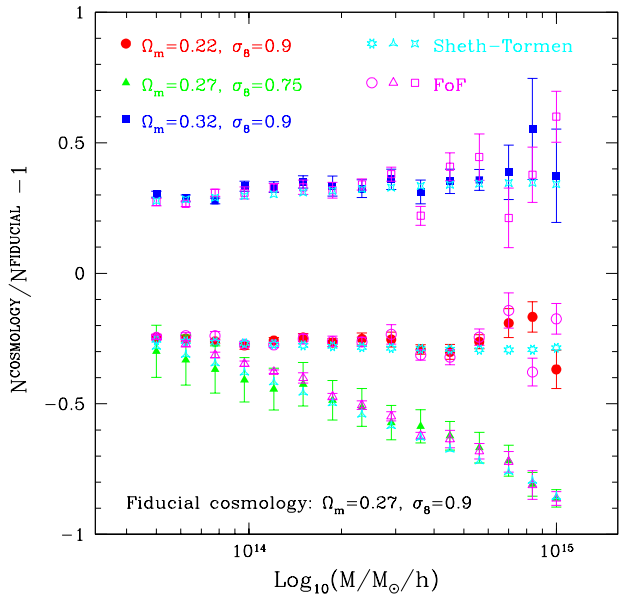}
\caption{The scaling with cosmology of the projected-peak mass
function. The mass function for each of the three cosmologies is scaled
by the fiducial cosmology mass function. The filled symbols represent
projected peaks, empty symbols the FoF halos, and starred ones the
ST theory. The error bars represent errors on the mean of the
eight realizations of each cosmology.}
\label{fig:mf_frac}
\end{figure}

However, our goal here is not to establish the correspondence between
the two-dimensional and three-dimensional masses of individual halos, but the cosmology
dependence of the shear-peak-abundance. This should be independent
of the halo mass definition and the choice of filter.

Figure~\ref{fig:mf_frac} is the most important result of this work. It
shows the mass function fractional difference of the variational
models and the fiducial one, averaged over the ensemble. The
fractional mass function for each model is depicted with a different
symbol--circle, triangle, rectangle; the filled symbols represent
projected-peaks measurements, FoF halos are the empty symbols, and
finally the ST theory is shown by the starred symbols. Just like in
Figure~\ref{fig:mf_abs}, the errors are on the mean of the eight
realizations of the three fractional differences.

The plot shows clearly that the projection mass function varies with
cosmology \mx{\it{in the same way}} as the three-dimensional and the ST mass
functions. Although this result was suggested by early studies
\cite{1999A&A...351..815R}, our simulations demonstrate with high
statistical significance that this conclusion is robust. There are two
reasons why this plot outlines unambiguously the scaling with
cosmology of the lensing mass function. First, the halo finder
differences that we have mentioned in the discussion of
Figure~\ref{fig:mf_abs} do not play a significant part here, since we
take ratios of numbers of objects selected with the same type of
filter. Second, we have reduced the cosmic variance of different
cosmologies, due to the way we chose the initial conditions for the
density fluctuations.

The most important consequence of the close cosmology scaling in
Figure~\ref{fig:mf_frac}, is that we can fit the projected mass
function for the fiducial model and then use the ST theory to predict
the behaviour for other cosmologies. We reserve a detailed
demonstration of this for our companion paper.

The results so far are very encouraging in demonstrating that the
projected mass function in 50--100 Mpc slabs has very similar
cosmological dependence to the better-studied virial mass functions.
Some caution is necessary, however, as shear maps are linear
combinations of projected density fields over many slabs along the
line of sight, weighted by the lensing kernel. Additional work is
needed when moving from the projected-density mass function to the
shear-peak distribution function. There is great hope that the
shear-peak mass function also follows the scaling with cosmology of
the three-dimensional mass function. Taking this step is our goal for the near
future. With the cosmological dependence of the projected mass
function well understood, WL-selected cluster catalogs can match the
statistical power of X-ray or Sunyaev-Zeldovich-selected catalogs of
similar depth, but sidestep the difficult issue of calibrating the
mass-observable relation.

\subsection*{Acknowledgements}
L.M. particularly thanks Tsz Yan Lam for numerous technical
discussions. We thank Ravi Sheth and Bhuvnesh Jain for access to their
Opteron cluster. We are also grateful to Peter Schneider for comments
on this manuscript. R.E.S. acknowledges support from a Marie Curie
Reintegration Grant and the Swiss National Foundation. G.B. and L.M. were
supported by NSF grant AST-0607667 and NASA grant
BEFS-04-0014-0018. G.B. additionally acknowledges Department of Energy
grant DOE-DE-FG02-95ER40893. L.M. is also supported by the Deutsche
Forschungsgemeinschaft under the Transregion TRR33 The Dark Universe.


\begin{thebibliography}{22}
\expandafter\ifx\csname natexlab\endcsname\relax\def\natexlab#1{#1}\fi
\expandafter\ifx\csname bibnamefont\endcsname\relax
  \def\bibnamefont#1{#1}\fi
\expandafter\ifx\csname bibfnamefont\endcsname\relax
  \def\bibfnamefont#1{#1}\fi
\expandafter\ifx\csname citenamefont\endcsname\relax
  \def\citenamefont#1{#1}\fi
\expandafter\ifx\csname url\endcsname\relax
  \def\url#1{\texttt{#1}}\fi
\expandafter\ifx\csname urlprefix\endcsname\relax\def\urlprefix{URL }\fi
\providecommand{\bibinfo}[2]{#2}
\providecommand{\eprint}[2][]{\url{#2}}

\bibitem[{\citenamefont{{Press} and {Schechter}}(1974)}]{1974ApJ...187..425P}
\bibinfo{author}{\bibfnamefont{W.~H.} \bibnamefont{{Press}}} \bibnamefont{and}
  \bibinfo{author}{\bibfnamefont{P.}~\bibnamefont{{Schechter}}},
  \bibinfo{journal}{\apj} \textbf{\bibinfo{volume}{187}}, \bibinfo{pages}{425}
  (\bibinfo{year}{1974}).

\bibitem[{\citenamefont{{Sheth} and {Tormen}}(1999)}]{1999MNRAS.308..119S}
\bibinfo{author}{\bibfnamefont{R.~K.} \bibnamefont{{Sheth}}} \bibnamefont{and}
  \bibinfo{author}{\bibfnamefont{G.}~\bibnamefont{{Tormen}}},
  \bibinfo{journal}{\mnras} \textbf{\bibinfo{volume}{308}},
  \bibinfo{pages}{119} (\bibinfo{year}{1999}).

\bibitem[{\citenamefont{{Warren} et~al.}(2006)\citenamefont{{Warren},
  {Abazajian}, {Holz}, and {Teodoro}}}]{2006ApJ...646..881W}
\bibinfo{author}{\bibfnamefont{M.~S.} \bibnamefont{{Warren}}},
  \bibinfo{author}{\bibfnamefont{K.}~\bibnamefont{{Abazajian}}},
  \bibinfo{author}{\bibfnamefont{D.~E.} \bibnamefont{{Holz}}},
  \bibnamefont{and}
  \bibinfo{author}{\bibfnamefont{L.}~\bibnamefont{{Teodoro}}},
  \bibinfo{journal}{\apj} \textbf{\bibinfo{volume}{646}}, \bibinfo{pages}{881}
  (\bibinfo{year}{2006}).

\bibitem[{\citenamefont{{Tinker} et~al.}(2008)\citenamefont{{Tinker},
  {Kravtsov}, {Klypin}, {Abazajian}, {Warren}, {Yepes}, {Gottl{\"o}ber}, and
  {Holz}}}]{2008ApJ...688..709T}
\bibinfo{author}{\bibfnamefont{J.}~\bibnamefont{{Tinker}}},
  \bibinfo{author}{\bibfnamefont{A.~V.} \bibnamefont{{Kravtsov}}},
  \bibinfo{author}{\bibfnamefont{A.}~\bibnamefont{{Klypin}}},
  \bibinfo{author}{\bibfnamefont{K.}~\bibnamefont{{Abazajian}}},
  \bibinfo{author}{\bibfnamefont{M.}~\bibnamefont{{Warren}}},
  \bibinfo{author}{\bibfnamefont{G.}~\bibnamefont{{Yepes}}},
  \bibinfo{author}{\bibfnamefont{S.}~\bibnamefont{{Gottl{\"o}ber}}},
  \bibnamefont{and} \bibinfo{author}{\bibfnamefont{D.~E.}
  \bibnamefont{{Holz}}}, \bibinfo{journal}{\apj}
  \textbf{\bibinfo{volume}{688}}, \bibinfo{pages}{709} (\bibinfo{year}{2008}),
  \eprint{0803.2706}.

\bibitem[{\citenamefont{{Marian} and {Bernstein}}(2006)}]{2006PhRvD..73l3525M}
\bibinfo{author}{\bibfnamefont{L.}~\bibnamefont{{Marian}}} \bibnamefont{and}
  \bibinfo{author}{\bibfnamefont{G.~M.} \bibnamefont{{Bernstein}}},
  \bibinfo{journal}{\prd} \textbf{\bibinfo{volume}{73}},
  \bibinfo{pages}{123525} (\bibinfo{year}{2006}).

\bibitem[{\citenamefont{{Wang} et~al.}(2004)\citenamefont{{Wang}, {Khoury},
  {Haiman}, and {May}}}]{2004PhRvD..70l3008W}
\bibinfo{author}{\bibfnamefont{S.}~\bibnamefont{{Wang}}},
  \bibinfo{author}{\bibfnamefont{J.}~\bibnamefont{{Khoury}}},
  \bibinfo{author}{\bibfnamefont{Z.}~\bibnamefont{{Haiman}}}, \bibnamefont{and}
  \bibinfo{author}{\bibfnamefont{M.}~\bibnamefont{{May}}},
  \bibinfo{journal}{\prd} \textbf{\bibinfo{volume}{70}},
  \bibinfo{pages}{123008} (\bibinfo{year}{2004}).

\bibitem[{\citenamefont{{Hoekstra}}(2001)}]{2001A&A...370..743H}
\bibinfo{author}{\bibfnamefont{H.}~\bibnamefont{{Hoekstra}}},
  \bibinfo{journal}{\aap} \textbf{\bibinfo{volume}{370}}, \bibinfo{pages}{743}
  (\bibinfo{year}{2001}).

\bibitem[{\citenamefont{{Hoekstra}}(2003)}]{2003MNRAS.339.1155H}
\bibinfo{author}{\bibfnamefont{H.}~\bibnamefont{{Hoekstra}}},
  \bibinfo{journal}{\mnras} \textbf{\bibinfo{volume}{339}},
  \bibinfo{pages}{1155} (\bibinfo{year}{2003}).

\bibitem[{\citenamefont{{Dodelson}}(2004)}]{2004PhRvD..70b3008D}
\bibinfo{author}{\bibfnamefont{S.}~\bibnamefont{{Dodelson}}},
  \bibinfo{journal}{\prd} \textbf{\bibinfo{volume}{70}},
  \bibinfo{pages}{023008} (\bibinfo{year}{2004}).

\bibitem[{\citenamefont{{Hamana} et~al.}(2004)\citenamefont{{Hamana}, {Takada},
  and {Yoshida}}}]{2004MNRAS.350..893H}
\bibinfo{author}{\bibfnamefont{T.}~\bibnamefont{{Hamana}}},
  \bibinfo{author}{\bibfnamefont{M.}~\bibnamefont{{Takada}}}, \bibnamefont{and}
  \bibinfo{author}{\bibfnamefont{N.}~\bibnamefont{{Yoshida}}},
  \bibinfo{journal}{\mnras} \textbf{\bibinfo{volume}{350}},
  \bibinfo{pages}{893} (\bibinfo{year}{2004}).

\bibitem[{\citenamefont{{Tang} and {Fan}}(2005)}]{2005ApJ...635...60T}
\bibinfo{author}{\bibfnamefont{J.~Y.} \bibnamefont{{Tang}}} \bibnamefont{and}
  \bibinfo{author}{\bibfnamefont{Z.~H.} \bibnamefont{{Fan}}},
  \bibinfo{journal}{\apj} \textbf{\bibinfo{volume}{635}}, \bibinfo{pages}{60}
  (\bibinfo{year}{2005}).

\bibitem[{\citenamefont{{De Putter} and {White}}(2005)}]{2005NewA...10..676D}
\bibinfo{author}{\bibfnamefont{R.}~\bibnamefont{{De Putter}}} \bibnamefont{and}
  \bibinfo{author}{\bibfnamefont{M.}~\bibnamefont{{White}}},
  \bibinfo{journal}{New Astronomy} \textbf{\bibinfo{volume}{10}},
  \bibinfo{pages}{676} (\bibinfo{year}{2005}).

\bibitem[{\citenamefont{{Hennawi} and {Spergel}}(2005)}]{2005ApJ...624...59H}
\bibinfo{author}{\bibfnamefont{J.~F.} \bibnamefont{{Hennawi}}}
  \bibnamefont{and} \bibinfo{author}{\bibfnamefont{D.~N.}
  \bibnamefont{{Spergel}}}, \bibinfo{journal}{\apj}
  \textbf{\bibinfo{volume}{624}}, \bibinfo{pages}{59} (\bibinfo{year}{2005}).

\bibitem[{\citenamefont{{Davis} et~al.}(1985)\citenamefont{{Davis},
  {Efstathiou}, {Frenk}, and {White}}}]{1985ApJ...292..371D}
\bibinfo{author}{\bibfnamefont{M.}~\bibnamefont{{Davis}}},
  \bibinfo{author}{\bibfnamefont{G.}~\bibnamefont{{Efstathiou}}},
  \bibinfo{author}{\bibfnamefont{C.~S.} \bibnamefont{{Frenk}}},
  \bibnamefont{and} \bibinfo{author}{\bibfnamefont{S.~D.~M.}
  \bibnamefont{{White}}}, \bibinfo{journal}{\apj}
  \textbf{\bibinfo{volume}{292}}, \bibinfo{pages}{371} (\bibinfo{year}{1985}).

\bibitem[{\citenamefont{{Kruse} and {Schneider}}(1999)}]{1999MNRAS.302..821K}
\bibinfo{author}{\bibfnamefont{G.}~\bibnamefont{{Kruse}}} \bibnamefont{and}
  \bibinfo{author}{\bibfnamefont{P.}~\bibnamefont{{Schneider}}},
  \bibinfo{journal}{\mnras} \textbf{\bibinfo{volume}{302}},
  \bibinfo{pages}{821} (\bibinfo{year}{1999}).

\bibitem[{\citenamefont{{Reblinsky} et~al.}(1999)\citenamefont{{Reblinsky},
  {Kruse}, {Jain}, and {Schneider}}}]{1999A&A...351..815R}
\bibinfo{author}{\bibfnamefont{K.}~\bibnamefont{{Reblinsky}}},
  \bibinfo{author}{\bibfnamefont{G.}~\bibnamefont{{Kruse}}},
  \bibinfo{author}{\bibfnamefont{B.}~\bibnamefont{{Jain}}}, \bibnamefont{and}
  \bibinfo{author}{\bibfnamefont{P.}~\bibnamefont{{Schneider}}},
  \bibinfo{journal}{\aap} \textbf{\bibinfo{volume}{351}}, \bibinfo{pages}{815}
  (\bibinfo{year}{1999}).

\bibitem[{\citenamefont{{Schneider}}(1996)}]{1996MNRAS.283..837S}
\bibinfo{author}{\bibfnamefont{P.}~\bibnamefont{{Schneider}}},
  \bibinfo{journal}{\mnras} \textbf{\bibinfo{volume}{283}},
  \bibinfo{pages}{837} (\bibinfo{year}{1996}).

\bibitem[{\citenamefont{{Springel}}(2005)}]{2005MNRAS.364.1105S}
\bibinfo{author}{\bibfnamefont{V.}~\bibnamefont{{Springel}}},
  \bibinfo{journal}{\mnras} \textbf{\bibinfo{volume}{364}},
  \bibinfo{pages}{1105} (\bibinfo{year}{2005}).

\bibitem[{\citenamefont{{Crocce} et~al.}(2006)\citenamefont{{Crocce},
  {Pueblas}, and {Scoccimarro}}}]{2006MNRAS.373..369C}
\bibinfo{author}{\bibfnamefont{M.}~\bibnamefont{{Crocce}}},
  \bibinfo{author}{\bibfnamefont{S.}~\bibnamefont{{Pueblas}}},
  \bibnamefont{and}
  \bibinfo{author}{\bibfnamefont{R.}~\bibnamefont{{Scoccimarro}}},
  \bibinfo{journal}{\mnras} \textbf{\bibinfo{volume}{373}},
  \bibinfo{pages}{369} (\bibinfo{year}{2006}).

\bibitem[{\citenamefont{{Marian} et~al.}()\citenamefont{{Marian}, {Smith}, and
  {Bernstein}}}]{2009M}
\bibinfo{author}{\bibfnamefont{L.}~\bibnamefont{{Marian}}},
  \bibinfo{author}{\bibfnamefont{R.~E.} \bibnamefont{{Smith}}},
  \bibnamefont{and} \bibinfo{author}{\bibfnamefont{G.~M.}
  \bibnamefont{{Bernstein}}} (In preparation).

\bibitem[{\citenamefont{{Metzler} et~al.}(2001)\citenamefont{{Metzler},
  {White}, and {Loken}}}]{2001ApJ...547..560M}
\bibinfo{author}{\bibfnamefont{C.~A.} \bibnamefont{{Metzler}}},
  \bibinfo{author}{\bibfnamefont{M.}~\bibnamefont{{White}}}, \bibnamefont{and}
  \bibinfo{author}{\bibfnamefont{C.}~\bibnamefont{{Loken}}},
  \bibinfo{journal}{\apj} \textbf{\bibinfo{volume}{547}}, \bibinfo{pages}{560}
  (\bibinfo{year}{2001}), \eprint{arXiv:astro-ph/0005442}.

\bibitem[{\citenamefont{{Clowe} et~al.}(2004)\citenamefont{{Clowe}, {De Lucia},
  and {King}}}]{2004MNRAS.350.1038C}
\bibinfo{author}{\bibfnamefont{D.}~\bibnamefont{{Clowe}}},
  \bibinfo{author}{\bibfnamefont{G.}~\bibnamefont{{De Lucia}}},
  \bibnamefont{and} \bibinfo{author}{\bibfnamefont{L.}~\bibnamefont{{King}}},
  \bibinfo{journal}{\mnras} \textbf{\bibinfo{volume}{350}},
  \bibinfo{pages}{1038} (\bibinfo{year}{2004}),
  \eprint{arXiv:astro-ph/0402302}.

\end{thebibliography}

\end{document}